\documentstyle[12pt]{article}
\begin{document}

\overfullrule 0 mm \language 0

\vskip 0.3 cm
\centerline {\bf{ RADIATION REACTION}}
\centerline {\bf{FOR A CHARGED BROWNIAN PARTICLE}}
 
\vskip 1.0 cm

\centerline {\bf{ Alexander A.  Vlasov}}
\vskip 0.3 cm
\centerline {{  High Energy and Quantum Theory}}
\centerline {{Department of Physics}}
\centerline {{ Moscow State University}}
\centerline {{  Moscow, 119899}}
\centerline {{ Russia}}
\vskip 0.3cm

\vskip 0.3 cm
{\it  As it is known a model of a charged particle with finite size is a 
good tool to consider the effects of self- action and backreaction, 
caused by electromagnetic radiation. In this work the "size" of a 
charged particle is induced by its stochastic Brownian vibration. 
Appropriate equation of particle's motion with radiation force is 
derived.   It is shown that the solutions of 
this equation correctly describe the effects of radiation reaction.} 
\vskip 1.0 cm

It is known from the classical electrodynamics that a charged particle, 
moving with acceleration, must radiate electromagnetic waves and thus 
must feel the backreaction of such radiation. How one can take into 
account this backreaction in the  equation of motion  of a 
particle ?

 This problem is very old ( 
count from the pioneer works of Abragam and Lorentz [1] ), but till know 
it is under focus of different investigations, handling it as from quantum 
field theory point of view, so  from point of view of the classical 
physics.

 The origin of the problem is the following. If in frames of some 
theoretical investigation the size of radiating body can be neglected 
( in comparison with other characteristic lengths),  then one can try to 
use the notion of "point-like" particle for such body and its mathematical 
representation - the Dirac's delta-function.  But  
the use of  the delta-function inevitably leads to divergences in 
some physical quantities, such as self- electromagnetic energy of point 
charged particle, its effective mass and so on, and consequently to the 
necessity of mass renormalization. Dirac was the first [2] to do this 
renormalization and as the 
result the famous equation with relativistic  radiation reaction for 
point-like particle was derived ( in the literature - the Abragam - 
Lorentz -Dirac equation (ALD) ).

But immediately scientists found out ( and Dirac was among them) 
that ALD equation leads to many paradoxes. 
Among them in the literature are usually mentioned [3] the following:

a) the existence of runaway solutions ( in the absences of external forces 
the radiating particle begins to move  with growing velocity up to that of 
light);

b) preacceleration (after supplementary condition, excluding the 
runaway solutions, the solutions remain, describing 
particle, "feeling" the external force with some advance in 
time);

c) the existence of "exotic" ALD solutions for head-on-collisions (i.e. 
for two opposite charged particles there are solutions, 
describing their mutual repelling);

and so on.

Thus one can understand the scientists, declaring that "the ALD equation 
must be modified".

There are two basic ways to do such modification:

a) to consider problem of "point-like" particle not in the frames of 
classical theory, but within quantum field theory, studying the processes 
of interaction of charged quantum particles with their quantum 
electromagnetic fields;

b) to stay in frames of classical theory, but refusing the 
notion of "point-like" particle and considering small object of finite 
size like charged particles of dusty plasma, Brownian charged particles 
and others.

In his works on radiation theme the author follows the second path, 
considering  nonquantum charged particles of small finite size [4].

In particular in this work is  investigated the influence of 
induced by Brownian vibration the effective  size of charged 
particle (in appropriate time scale) on the effect of radiation reaction. 
The equation of center mass motion of such particle is derived. Some 
solutions of this equation are investigated.

At first let us remind that the explicit expression of electric field 
$\vec E$ for some moving charged body, $$\vec E = -\nabla \phi -{1\over 
c}\cdot {\partial \vec A \over \partial t},$$

taking into account the effects of  retardation: 

$$\phi(t,\vec r) = \int dt' d\vec r' \delta(t'-t+|\vec r  -\vec 
r'|/c)\cdot {1 \over |\vec r  -\vec r'|}\cdot \rho(t',\vec r');$$
$$\vec A(t,\vec r) = \int dt' d\vec r' \delta(t'-t+|\vec r  -\vec 
r'|/c)\cdot {1 \over c |\vec r  -\vec r'|}\cdot \vec j (t',\vec r')$$

and with the help of the charge conservation law:
$${\ \partial \rho \over \partial t}+ div \ \vec j =0,$$
can be expanded in series in retardation [5]

$$\vec E(t, \vec 
r)=-\int  d\vec r' \ \rho' \nabla |\vec r-\vec r'|^{-1} + $$ 
 
 $$+ \sum_{m=1}^{\infty} {(-1)^m \over c^m m!}\cdot{\partial ^{m-1} 
\over \partial t^{m-1}} \int  d\vec r' \ (\vec j', \nabla) \nabla 
 |\vec r - \vec r'|^{m-1}-$$ $$- \sum_{m=0}^{\infty} {(-1)^m \over c^{m+2} 
 m!}\cdot{\partial ^{m+1} \over \partial t^{m+1}}\int  d\vec r' \ \vec j' \ 
 |\vec r - \vec r'|^{m-1}\eqno(1)$$
 here $\vec j' \equiv \vec j (t, \vec r')$.

The   average of  expression (1) over the body's volume:
 $$\vec E \to <\vec E>,$$ 
 $$ <\vec E> \equiv \int d\vec r \rho(t,\vec r) \vec E (t,\vec r)/ 
\bigg(\int d\vec r \rho(t,\vec r)\bigg) \eqno (2)$$ gives us the 
average  self- electric field of a moving body.

For spherically symmetric charge distribution:  
$$\rho = \rho(t,|\vec r - \vec R (t)|)$$ 
the space derivatives in (2) and (1) are averaged according to 
the rule: $$<\nabla_{\alpha}\nabla_{\beta}>= 
\delta_{\alpha\ \beta} \nabla^2 /3,\ \ <\nabla> =0.$$ Then in (2) 
the first (Coulomb) term of the expression (1) vanishes, 
and the remaining two sums   can be reduced  in such a way that $$<\vec 
E> =-{2 \over 3Q c^2} \sum_{n=0}^{\infty} {(-1)^n \over c^{n} n!}\int 
d\vec r \rho(t,\vec r) \cdot{\partial ^{n+1} \over \partial t^{n+1}}\int 
d\vec r' \ \vec j' \ |\vec r - \vec r'|^{n-1}$$ 
This expression can be simplified further:
  $$ <\vec E> =-{2 \over 3 Qc^2} \int 
d\vec r \rho(t,\vec r) \cdot{\partial \over \partial t} \int {d\vec r' 
\over |\vec r - \vec r'|} \sum_{n=0}^{\infty} \bigg(- {|\vec r - \vec r'| 
\over c} \cdot{\partial \over \partial t}\bigg)^{n} {1 \over n!}\ \  \vec 
j'= $$ 
$$ =-{2 
\over 3Qc^2} \int d\vec r \rho(t,\vec r) \cdot{\partial \over \partial t} 
 \int {d\vec r' \over |\vec r - \vec r'|} \vec j(t-{|\vec r - \vec r'| 
\over c},\  \vec r') \eqno (3)$$ It should be mentioned that the 
 expression (3) is a strict one for spherically symmetric charged 
body.

Let us made new simplifications.

Let the body be rigid in the sense that $$\vec j(t_{ret}, \vec r') = 
\rho(t_{ret}, \vec r') \cdot \vec v (t_{ret}, \vec r') \eqno (4)$$ here 
$t_{ret}\equiv t-{|\vec r - \vec r'| \over c}$ - the retarded time.

 Then let us consider the retardation only for the velocity $v$ in (4) 
and neglect the retardation for the density of charge:  $$\rho(t_{ret}, 
  \vec r')\approx \rho(t, \vec r') \eqno(5)$$ This means that Taylor 
expansion in powers of $c\to \infty$ (expansion in retardation) for the 
velocity is more sufficient then for the density of charge, i.e.  $$v 
{\partial \over \partial t} \rho \ll \rho {\partial \over \partial t}v 
\sim \rho F_{ext}/m \eqno (6)$$ If $T_{\rho}$ - is the typical time of 
density variation, and $T_{v}$ - is the typical time of velocity 
variation, then this inequality can be rewritten as:  $$T_{v} \ll 
T_{\rho}$$ The inequality (6) has one more interpretation.  Due to the law 
 of charge conservation:   $${\partial \over \partial t} \rho= 
 -\bigg(\nabla,\ \ \vec j \bigg)\sim \rho v$$ inequality (6) leads us to 
 the linearity condition, when one consider only terms linear in velocity and 
its time derivatives terms.

Thus in (3) following (4, 5, 6) one can apply the time derivative only 
to the velocity (and not to the density). This leads to the following 
expression for average electric field:  $$<\vec E> 
=-{2 \over 3Qc^2} \int d\vec r' d\vec r \cdot {\rho(t,\vec r) \rho(t,\vec 
r') \over |\vec r - \vec r'|} {\partial \over \partial t}\ \vec v(t-{|\vec 
r - \vec r'| \over c}) \eqno (7)$$ Expression (7), multiplied on the value 
of the charge $Q$, is just the Jackson self- electromagnetic force [5] in 
linear approximation (the magnetic Lorentz self- force is zero in 
approximation under consideration).

Let us note that in (7) one can consider varying in time densities if 
the inequality (6) is valid.

\vskip 0.2 cm

Now  turn to the motion of charged Brownian particle.

  Let us take those time intervals $T_{v}$ (time scales), for which the  Brownian 
motion can be described by distribution function: $$ T_{v}\gg 
T_{Br}$$
here $T_{Br}$ - is the typical time for Maxwell's velocity distribution to 
appear.

Let $n(t, \vec r)$   be the probability to find Brownian 
particle at the moment of time $t$ in the volume $d \vec r$ 
, thus $n$ - is the distribution function with norm 
$$\int n(t, \vec r) 
d\vec r =1$$ 
This function obeys  the conservation law $${\partial n \over \partial t 
}+ div\  n \vec V =0 \eqno (8)$$ 

Let the  motion of the Brownian particle consists from two  motions -
the first one, regular, with the velocity 
$\vec v= \vec v (t)$, under the influence of some external regular 
force 
$\vec F_{ext}$, and the second one, irregular - Brownian diffusion with 
velocity 
 $\vec u$ :  $$\vec V =\vec v +\vec u,$$ $$ \vec u 
=-{D \over n} \nabla n \eqno (9)$$ here $D$ - the diffusion 
parameter.

In other words around the regular trajectory of particle's center of 
mass there are Brownian vibration in such a way that the average 
value of the square particle  displacement from the regular trajectory 
is proportional, following Einstein formula, to
 $Dt $. Thus appears the effective particle's "size", proportional to  
$\sqrt{Dt}$.

Equation (8) with the help of (9) 
can be put in the form $${\partial n \over \partial t }+ 
(\vec v, \nabla) \ n =D\nabla^2 n \eqno (10)$$ As the particle's center 
of mass moves along its trajectory with velocity $\vec 
v =d \vec R (t) /dt$, the distribution function gives the probability of 
particle's displacement from this trajectory, so $$n =n(t, \vec r - \vec R 
(t) )$$ 
Inserting this form of $n$ into (10) and taking into account that
 $${\partial n \over \partial t }= -({d \vec 
R (t) \over dt},\ \nabla) n + \bigg({\partial n \over \partial t 
}\bigg)_{\vec R =const}$$ one gets  equation for 
$n$:  
$$\bigg({\partial n \over \partial t }\bigg)_{\vec R =const}=D \nabla^2 n 
\eqno (11)$$ It is the  typical Fokker - Plank equation with solution, for 
the initial condition $$n(0, \vec r) = \delta(\vec r- \vec R (0)),$$ in 
the form $$n(t, \vec r) ={1 \over (4\pi D t)^{3/2}}\exp{\bigg(- {|\vec 
r- \vec R (t)|^2 \over 4Dt}\bigg)}\eqno (12)$$ (see., for ex., [6]).

 To take into consideration the radiation reaction one must the 
Newtonian equation of the center of mass motion: 
 $$\vec a (t)={ \vec {F}_{ext}(t) \over m}$$ (here 
$\vec a(t) = \ddot {\vec R} (t)$) supplement with the self- force $\vec 
F_{self}= Q <\vec E>$ (see formula (7), where the spherically symmetric 
density of charge is $\rho =Q \cdot n$). 

Then the equation of motion of the center of mass in 
nonrelativistic approximation will be

$$\vec a (t)=
{ \vec {F}_{ext}(t) \over m}- {2 Q^2\over 3 m c^2 } \int \int d\vec r 
d\vec r'{n(t, \vec r)\ n(t, \vec r') \over |\vec r - \vec r'|} \vec a (t 
-{ |\vec r - \vec r'| \over c})\eqno(13)$$  

Using the result (12) for distribution 
function, introducing new dimensionless variables $$\vec r- \vec R (t) 
\equiv \vec \mu;\ \ \ |\vec \mu|\equiv \sqrt{4Dt} \ x;$$ $$\vec r'- \vec R 
(t) \equiv \vec \nu;\ \ \ |\vec \nu|\equiv \sqrt{4Dt} \ y;$$ and taking 
 into consideration the spherical symmetry, equation (13) finally reduces 
to this form $$\vec a (t)= { \vec {F}_{ext}(t) \over m}- $$ $$ - {4 
Q^2\over 3\pi m c }\cdot {1 \over Dt} 
\int\limits_{0}^{\infty}\int\limits_{0}^{\infty} dx dy\  xy \ 
e^{(-x^2-y^2)}\  \bigg[\vec v (t -{ \sqrt{4Dt} \over c} |x-y|)- \vec v 
(t -{ \sqrt{4Dt} \over c} |x+y|) \bigg] \eqno(14)$$

Thus equation (13) ( or (14) ) take into account the existence of 
 the finite size of the charged  particle, induced by Brownian vibration 
 (in time scale $T_{Br}\ll T_{v} \ll T_{\rho}$ ). Due to the finiteness 
  of particle's size, the  self-force, which is the radiation reaction 
 force in our approach, has finite value and there is no need to do mass 
 renormalization.
 
 Equation (13) differs from ALD equation, but has much in 
 common with the Sommerfeld finite-size models of a charged particle.

Let us mention the following features of equation (13) ((14)):

1). If the external force is constant:
 $\vec F_{ext} =\vec F_{0} =const$ then eq. (13) has the solution 
$$   \vec a  = {\vec {F}_0 \over m}\left(1+ {2Q^2 \over 3 m c^2 } \int 
\int d\vec r d\vec r'{n(t, \vec r)\ n(t, \vec r') \over |\vec r - \vec 
r'|}\right)^{-1} \equiv {\vec {F}_0 \over m+m_{em}^{*} } \eqno(15)$$

here $m_{em}^{*}$  - is the effective mass of the self- electromagnetic 
 field:  $$m_{em}^{*}  
\equiv {2Q^2 \over 3  c^2 } \int \int d\vec r d\vec r'{n(t, \vec r)\ n(t, 
\vec r') \over |\vec r - \vec r'|}$$
 
 If   the effective mass is approximated as ${Q^2 \over c^2 L_{Br}}$, 
where $ L_{Br}$ - the typical "size" of Brownian spread, equal according 
to Einstein formula, to $ \sqrt{4Dt}$, then, following (15), the 
acceleration of Brownian particle slightly increases in time up to its 
maximum value ${\vec {F}_0 \over m},$ achieved at $t \to \infty$. In other 
words, in this process the self-force (force of radiation reaction) is not 
equal to zero and together with the effective self- electromagnetic mass 
vanishes at $t \to \infty$.

2). If the external force is absent, one can find by ordinary 
substitution that eq. (14) has no "free" harmonic solutions like
 $$\vec v = \vec A \ cos {(wt)}$$ 
That is, contrary to Sommerfeld models (see, for ex., [4] )
there are no oscillations, free of radiation damping.

3). In details, "free" solutions of eq. (14)  are exponentially damped. 
Indeed, after substitution in (14) the velocity in the form $$\vec v \sim 
e^{(pt)},\ \ p=p' + i p''$$ where the real part of parameter $p$ is 
small enough:  $ p' \to 0\ (p' T_{v} \ll 1)$, one gets the following 
algebraic equation:
 $$p = - {4 
Q^2\over 3\pi m c }\cdot {1 \over Dt} 
\int\limits_{0}^{\infty}\int\limits_{0}^{\infty} dx dy\  xy \ 
e^{(-x^2-y^2)}\  \bigg[e^{(-\delta p |x-y|)}- e^{(-\delta 
 p |x+y|)} \bigg]  $$ here $\delta \equiv { \sqrt{4Dt}  \over c} $.
 
The real part of it provides us with
this equation
  $$p'=$$
$$=- {4 Q^2\over 3\pi m c }\cdot {1 \over Dt} 
 \int\limits_{0}^{\infty}\int\limits_{0}^{\infty} dx dy\  xy \ 
e^{(-x^2-y^2)}\ \cdot$$
$$\cdot \bigg[e^{(-\delta p'|x-y|)}cos (\delta p''
 |x-y|)- e^{(-\delta p'|x+y|)}cos (\delta p''|x+y|)  \bigg]  \approx$$

 $$\approx - {4 Q^2\over 3\pi m c }\cdot {1 \over Dt} 
\int\limits_{0}^{\infty}\int\limits_{0}^{\infty} dx dy\  xy \ 
 e^{(-x^2-y^2 )}\  \bigg[cos 
 (\delta p''|x-y|)- cos 
 (\delta p''|x+y|)  \bigg]  =$$
 $$=- {4 Q^2\over 3\pi m c }\cdot {1 \over Dt} 
\bigg(\int\limits_{0}^{\infty} dx \  x \ 
 e^{(-x^2 )}sin (\delta p''x) \bigg)^2 <0$$
 Consequently
 $$p'<0.$$
 This means that the solutions of (14) for  zero external force are 
exponentially damped.

 Such damping is obvious from the physical point of view - it is 
caused by the radiation energy losses.

 4). Let us expand the integrand in (14) in series in "retardation":  $$t 
\gg { \sqrt{4Dt} \over c} |x\pm y|$$ Then, taking into account $$ \vec v (t - \delta |x-y|)- \vec v (t -\delta |x+y|) 
\approx \dot {\vec v} (t) (x+y -|x-y|) \delta -2xy \delta ^2 \ddot {\vec 
v} (t) + ...$$ after integration one gets $$\vec F_{self}/m 
\approx -\dot {\vec v} (t){ m_{em} \over m} +{2Q^2 \over 3 m c^3} \ddot 
{\vec v} (t) +...  \eqno (16)$$ here the second term is the classical 
Abragam - Lorentz expression for radiation force and the effective 
electromagnetic mass $m_{em}$ of the first term 
equals to $$ {Q^2 \over c^2} \cdot {1 \over \sqrt{Dt} } \cdot {\sqrt 
{2} \over 3 \sqrt{\pi} } \eqno (17)$$ and tends to zero for $t \to \infty$.

Thus we have found out that the consideration of the Brownian spread of 
particle's size (in appropriate time scale) leads to new equation of 
particle's  motion with solutions having explicit physical sense and   
avoids the troubles connected with the notion "point-like" particle.

It should be noted that the idea to consider the existence of particle's 
 size, induced by Brownian vibration, is not the original one (see, 
for ex.,  the discussion and references in the chapter 22 of the book 
[8]).  Nevertheless author does not know the works where this idea was 
realized in concrete mathematical equations.  In our work - this is the 
equation (13). It is integro - differential - difference equation with 
retardation.  That is why it may have solutions, besides mentioned above, 
typical for the models of particles of a finite size - the so called 
tunneling solutions [4].  It would be interesting to investigate this 
 problem more closely in the further works.

\eject

 \centerline {\bf{REFERENCES}}

  \begin{enumerate}
\item
H. Lorentz, "The Theory of Electron", Leipzig, Teubner, 2nd edition, 1916. 
M.Abragam," Electromagnetische Theorie der Strahlung", Leipzig, Teubner, 
1905.
\item P.Dirac, Proc. Roy. Soc., A167, 148, 1938. 

\item
 N.P.Klepikov, Usp. Fiz. Nauk, 146, 317 (1985).
 S.Parrott, {\it Relativistic Electrodynamics and Differential
Geometry}, Springer-Verlag, NY, 1987.
   T.Erber, Fortschr.Phys., 9, 342 (1961).
 P.Pearle, in {\it Electromagnetism}, ed. D.Tepliz, Plenum, NY,
1982, p.211.
 A.Yaghjian, {\it Relativistic Dynamics of a Charged Sphere},
 Lecture Notes in Physics, 11, Springer, Berlin, 1992.

\item Al.€.Vlasov, Vestnik Mosk.St.Univ., Fizika, N 5, 17 
(1998); N 6, 15 (2001).  Alexander A.Vlasov, in "Photon: old problems in 
light of new ideas", p.  126, ed. Valeri V. Dvoeglazov, Nova Sci. Publ., 
 NY, 2000.  E-print Archive: physics/9911059, physics/9912051, 
 physics/0004026, physics/0103065, physics/011003, physics/0205012. 

 \item J.D. Jackson {\it Classical Electrodynamics 
  }, Wiley, NY, 1999.

\item I.A.Kvasnikov {\it Thermodynamics and Statistical 
 Physics. Part 2.},  Moscow, 
 Mosk.St.Univ., 1988.

\item A.Sommerfeld, Gottingen Nachrichten, 29 (1904), 363 (1904), 201 
  (1905).
\item  €.€.Sokolov, Yu.M.Loskutov, I.M.Ternov {\it Quantum 
Mechanics}., Moscow, Uchpedgiz, 1962.
 
   \end{enumerate}

\end{document}